\documentclass{appolb}
\usepackage{graphicx}
\usepackage{hyperref}
\usepackage{amsmath}
\usepackage{amssymb}
\usepackage[numbers,sort&compress]{natbib}

\begin{document}

\title{Unquenching the three-gluon vertex: A status report
\thanks{Presented at Excited QCD 2015, 8-14 March 2015, Tatranska Lomnica, Slovakia}
}
\author{Adrian Lorenz Blum, Reinhard Alkofer, Markus Q. Huber
\address{Institute of Physics, University of Graz, NAWI Graz, Universit\"atsplatz 5, 8010 Graz, Austria}
\\ \vspace*{0.5cm}
{Andreas Windisch
}
\address{Department of Physics, Washington University, St. Louis, MO, 63130, USA}
}

\maketitle

\begin{abstract}
We discuss unquenching of the three-gluon vertex via its Dyson-Schwin\-ger equation. We review the role of Furry's theorem and present first results for the quark triangle diagrams using non-perturbatively calculated dressing functions for the quark propagator and the quark-gluon vertex.
\end{abstract}
  
\section{Introduction}

The three-gluon vertex was recently intensively investigated in pure Yang-Mills theory \cite{Pelaez:2013cpa,Aguilar:2013vaa,Blum:2014gna,Eichmann:2014xya}. Its quantitative influence on propagators was studied, for example, in \cite{Blum:2014gna}, but also for the quark-gluon vertex it plays an important role \cite{Windisch:2014th,Hopfer:2014th}. Consequently, the inclusion of unquenching effects constitutes a necessary step for future studies relying on the three-gluon vertex. Furthermore, the three-gluon vertex also enters in some beyond rainbow-ladder truncations of bound state equations, see, for instance, \cite{Fischer:2009jm}.

In the following we will give a status report on ongoing efforts of including quark effects into the three-gluon vertex via its Dyson-Schwinger equation (DSE). Essential for these efforts is thereby non-perturbative input for the quark propagator and the quark-gluon vertex, which was studied, for example, in  \cite{Alkofer:2008tt,Rojas:2013tza,Williams:2014iea,Aguilar:2014lha,Windisch:2014th,Hopfer:2014th,Mitter:2014wpa,Pelaez:2015tba}.

\section{Three-gluon vertex Dyson-Schwinger equation}

\begin{figure}[tb]
	\centerline{
		\includegraphics[width=12.5cm]{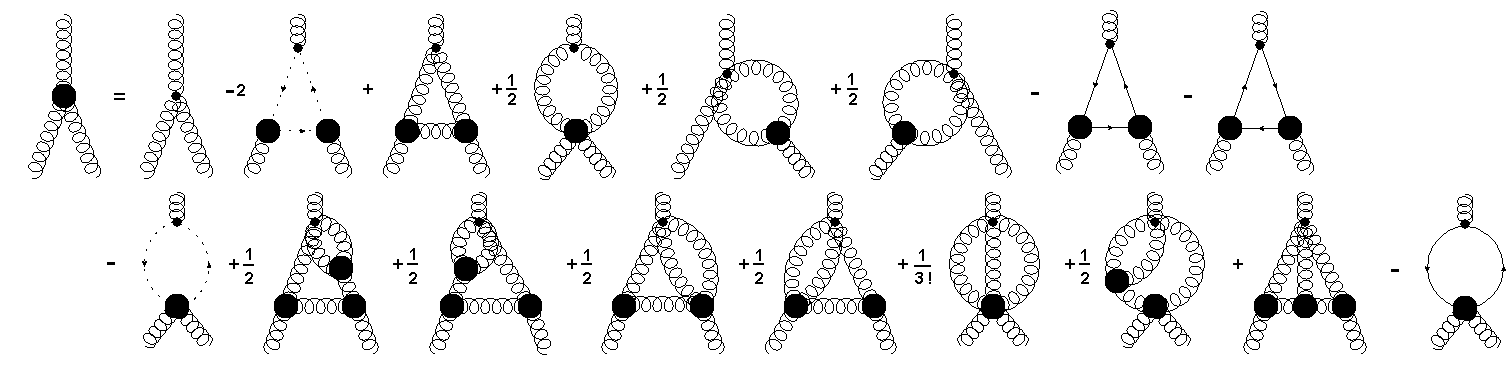}}
	\caption{The full DSE for the three-gluon vertex. Quarks enter via the quark triangle diagrams (7th and 8th diagrams on the right-hand side) and the quark swordfish diagram (last diagram). Thick/thin blobs denote dressed/bare vertices. All propagators are dressed. Wiggly/dashed/continuous lines represent gluons/ghosts/quarks.}
	\label{fig:unquenchedthreegluonvertex_all02}
\end{figure}

\begin{figure}[tb]
	\centerline{
		\includegraphics[width=12.5cm]{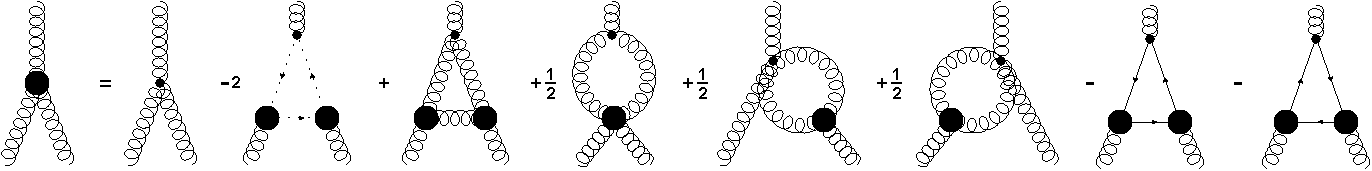}}
	\caption{The truncated DSE for the unquenched three-gluon vertex.}
	\label{fig:threegluonvertexunquenched02}
\end{figure}

The full DSE for the three-gluon vertex is depicted in Fig. \ref{fig:unquenchedthreegluonvertex_all02}. The diagrams with quarks are called the quark triangle and the quark swordfish diagrams.
To decouple the three-gluon vertex DSE from the equations of higher n-point functions (except for the four-gluon vertex) the following truncation is employed: All two-loop diagrams and diagrams with non-primitively divergent vertices are neglected. This truncation contains then all UV leading diagrams. The truncated equation is shown in Fig. \ref{fig:threegluonvertexunquenched02}. The only remaining higher n-point function is the four-gluon vertex, for which a model is employed. In Yang-Mills theory this truncation scheme turned out to be very successful \cite{Blum:2014gna,Eichmann:2014xya}. The qualitative appropriateness of the four-gluon vertex model was also confirmed \textit{a posteriori}, see \cite{Binosi:2014kka,Cyrol:2014kca} . However, within this scheme the quark swordfish diagram is dropped, which might yield important contributions \cite{Mitter:2014wpa,Hopfer:2014pc}. It remains for future investigations to include this diagram by using an appropriate model for the two-quark-two-gluon vertex.              

The pure Yang-Mills part of the three-gluon vertex DSE was solved self-consistently and with full momentum dependence in \cite{Blum:2014gna,Eichmann:2014xya} within the truncation described above.
One of the major findings was that the tree-level structure is dominant and the other three transverse dressing functions are subleading \cite{Eichmann:2014xya}. Therefore, we will consider only the tree-level structure as well.
In the following we will investigate the contributions from the quark triangles using results for the quark propagator and the quark-gluon vertex obtained elsewhere.

\subsection{Quark propagator and quark-gluon vertex}

For the quark-gluon vertex a 3PI equation is used instead of its DSE. This avoids the appearance of the unknown quark-gluon four-point function. 
The quark-gluon vertex can be decomposed into eight transverse and four longitudinal basis tensors. For Landau gauge, we only need the transverse part and we decompose the vertex as follows, see \cite{Windisch:2014th} for details:
\begin{align}
 \Gamma_\mu(p,q,p\cdot q)=\sum_{i=1}^{8}g_i(p,q,p\cdot q) \rho_\mu^i.
\end{align}
The tensors $\rho_\mu^i$, $i=1,4,6,7$, are chirally symmetric and those with $i=2,3,5,8$ are chirally anti-symmetric.

\begin{figure}[tb]
\centerline{
\includegraphics[width=8cm]{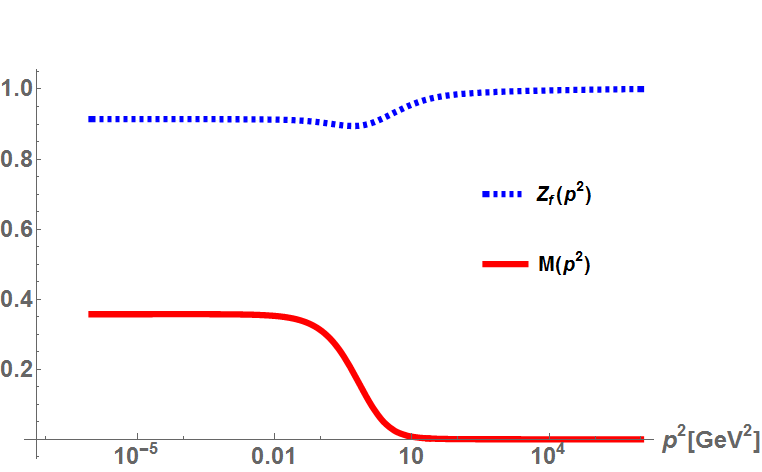}}
\caption{The quark wave function renormalization $Z_{f}(p^{2})$ and the quark mass function $M(p^{2})$ of the quark propagator.}
\label{Fig:qprop}
\end{figure}

\begin{figure}[bt]
\center
\begin{minipage}[bt]{6cm}
\center
		\includegraphics[width=6.00cm]{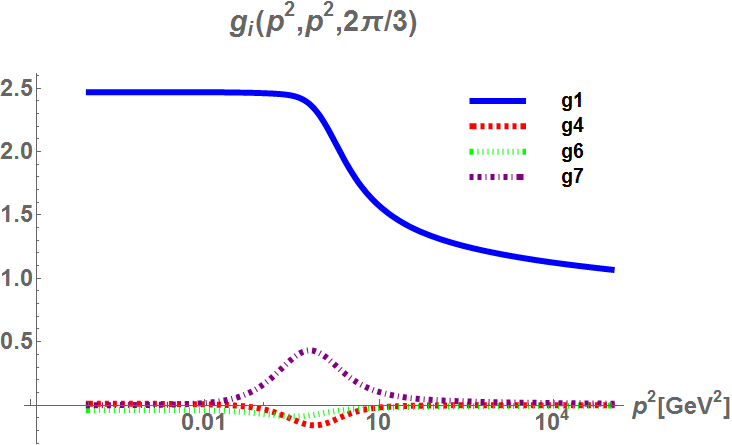}
\end{minipage}
\begin{minipage}[bt]{6cm}
\center
		\includegraphics[width=6.00cm]{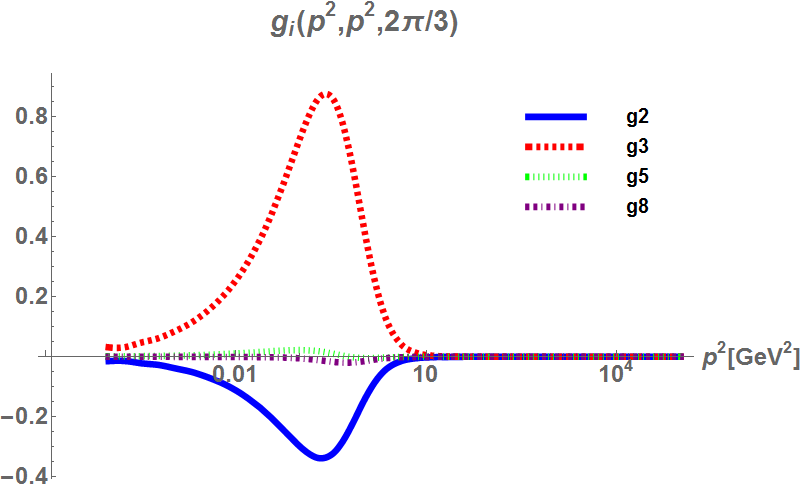}
\end{minipage}
\caption{The dressing functions of the quark-gluon vertex at the symmetric point corresponding to the eight transverse tensor structures. The left plot shows the results for the chirally symmetric and the right plot for the chirally anti-symmetric structures.}    
\label{dressing_qgv}
\end{figure}

The 3PI equation for the vertex is solved together with the quark propagator DSE in the chiral limit. A gluon propagator based on \cite{Huber:2012kd,Huber:2014tva} is used as an input. Results for the quark propagator, characterized by the quark wave function renormalization $Z_{f}(p^{2})$ and the quark mass function $M(p^{2})$, are shown in Fig.~\ref{Fig:qprop} and selected results for the vertex in Fig.~\ref{dressing_qgv}. Note that two of the largest dressing functions are $g_2$ and $g_3$, which are only allowed due to the breaking of the chiral symmetry.

\subsection{The quark triangles}

There are two quark triangle diagrams with opposite direction of momentum flow of the quarks, see Fig.~\ref{fig:threegluonvertexunquenched02}. This can be interpreted as quarks going round in one diagram and anti-quarks in the other. In QED similar diagrams exist for the three-photon vertex. However, according to Furry's theorem these two diagrams cancel each other. The argument can be generalized to any diagram with a closed fermion loop. The general result is that there is no photon vertex with an odd number of legs. The case of QED must be adapted to the non-Abelian character of QCD, viz., color has to be taken into account. The color part can be factorized and reads $\Tr T^a T^b T^c$ for one diagram, $\Tr T^a T^c T^b$ for the other, where $T^a$ are the generators of $SU(3)$. The color trace yields
\begin{align}
 \Tr T^a T^b T^c = \frac{1}{4}(d^{abc}+i\,f^{abc}).
\end{align}
In QED the two diagrams cancel, because they are exactly the same except for the signs. For QCD, however, the effect of the opposite signs is different, since the color parts of the two diagrams have a symmetric and an anti-symmetric part. The symmetric parts then cancel, as in QED, while the anti-symmetric parts add up. We thus need to calculate only one diagram, multiply it by a factor of $2$ and take only the anti-symmetric color part. The fact that $d^{abc}$ drops is also in agreement with \cite{Smolyakov:1980wq} where it was shown that the three-gluon vertex must be proportional to $f^{abc}$. Hence, we will consider only one quark triangle in the following.

\begin{figure}[tb]
\centerline{
\includegraphics[width=0.45\textwidth]{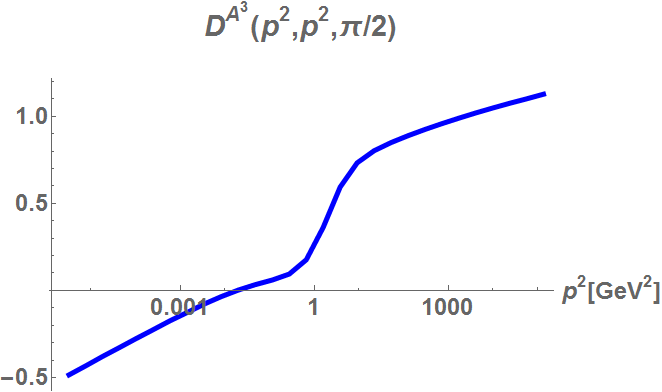}\hfill
\includegraphics[width=0.45\textwidth]{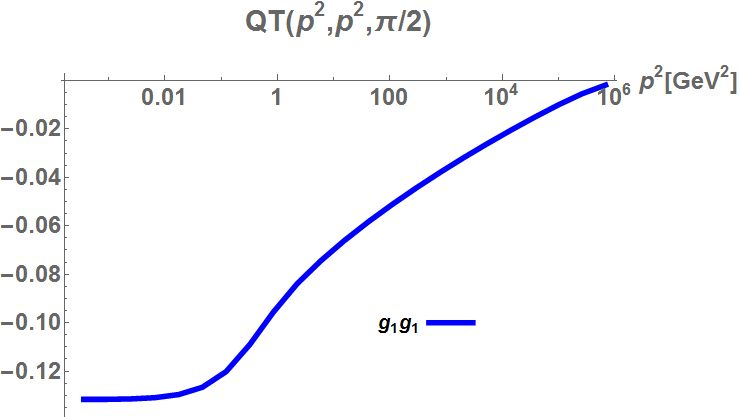}}
\caption{\textit{Left}: Three-gluon vertex result for the Yang-Mills part from \cite{Blum:2014gna}. \textit{Right}: The quark triangle diagram when only the tree-level dressing of the quark-gluon vertex, $g_1$, is considered.}
\label{Fig:F2H}
\end{figure}

The quark triangle diagram is a static diagram in the sense that it does not depend on the three-gluon vertex. Thus it needs to be calculated only once and its contribution can be added to the Yang-Mills part during the iteration process. This is advantageous, as there are 64 different combinations of dressing functions for the quark triangle diagram from the two dressed quark-gluon vertices with 8 dressings each. As an example we show the contribution calculated from the dressing $g_1$, corresponding to the transversely projected tree-level tensor, in Fig. \ref{Fig:F2H}. This particular contribution turns out to be rather small when compared to the full three-gluon vertex of pure Yang-Mills theory, see also Fig.~\ref{Fig:F2H}. Nevertheless, it can have quite some impact on the three-gluon vertex due to its zero crossing: Even a small change can move the zero crossing. Whether this is indeed the case will be investigated in future calculations. In particular the contributions from the other quark-gluon vertex dressing functions must be included to arrive at a final conclusion.

\section{Conclusions and outlook}

The three-gluon vertex plays an essential role for future DSE calculations in full QCD. Consequently, the inclusion of quark effects is an important extension of existing pure Yang-Mills calculations. We have set up a straightforward scheme that will allow unquenching of the vertex via its DSE. Essential input is thereby provided by non-perturbative results from the matter part of QCD. However, back-coupling effects of the three-gluon vertex onto the matter part are not included in this scheme.

\section*{Acknowledgments} 

Funding by the FWF (Austrian science fund) under Contract P 27380-N27 and through the Doctoral Program ``Hadrons in Vacuum, Nuclei and Stars'', Contract W1203-N16, is gratefully acknowledged.

\bibliographystyle{utphys_mod}
\bibliography{literature_eQCD2015_quarkTriangle}

\end{document}